\documentclass[runningheads]{llncs}
\usepackage{graphicx}
\usepackage{epstopdf}
\usepackage{bbding}
\usepackage{amsmath}
\usepackage{subfigure}
\usepackage{algorithm}
\usepackage{algorithmic}
\usepackage{longtable}
\usepackage{rotating}
\usepackage{multirow}
\usepackage{multicol}
\usepackage{tabularx}
\usepackage{booktabs}
\usepackage{threeparttable}

\begin{document}

\title{A Training-based Identification Approach to VIN Adversarial Examples}
%
%
\author{Yingdi Wang\inst{1} \and
Wenjia Niu\inst{ 1(}\Envelope\inst{) } \and
Tong Chen\inst{1}\and
Yingxiao Xiang\inst{1}\and
Jingjing Liu\inst{1}\and
Gang Li\inst{2}\and
Jiqiang Liu\inst{1}}
\authorrunning{Y. Wang et al.}
%
\institute{{Beijing Key Laboratory of Security and Privacy in Intelligent Transportation,\\
Beijing Jiaotong University, Beijing 100044, China}\and
{School of Information Technology, Deakin University}\\
\email{niuwj@bjtu.edu.cn}}
%
\maketitle              
\begin{abstract}
With the rapid development of Artificial Intelligence (AI), the problem of AI security has gradually emerged. Most existing machine learning algorithms may be attacked by adversarial examples. An adversarial example is a slightly modified input sample that can lead to a false result of machine learning algorithms. The adversarial examples pose a potential security threat for many AI application areas, especially in the domain of robot path planning. In this field, the adversarial examples obstruct the algorithm by adding obstacles to the normal maps, resulting in multiple effects on the predicted path. However, there is no suitable approach to automatically identify them. To our knowledge, all previous work uses manual observation method to estimate the attack results of adversarial maps, which is time-consuming. Aiming at the existing problem, this paper explores a method to automatically identify the adversarial examples in Value Iteration Networks (VIN), which has a strong generalization ability. We analyze the possible scenarios caused by the adversarial maps. We propose a training-based identification approach to VIN adversarial examples by combing the path feature comparison and path image classification. We evaluate our method using the adversarial maps dataset,  show that our method can achieve a high-accuracy and faster identification than manual observation method.
\end{abstract}
\section{Introduction}
Due to the advances in machine learning and deep learning algorithms, AI has developed rapidly in recent years. AI has helped humans solve many complicated problems, such as image classification, face recognition, robot path planning and so on\cite{russakovsky2015imagenet,schroff2015facenet,zhu2017target}.

However, with the development of AI, the vulnerability of the machine learning algorithms is gradually emerging. An adversary can take advantage of the characteristics of the algorithm to disguise itself in order to fool the detection of the classifier\cite{dalvi2004adversarial}. More typically, for deep neural network structures, a well-designed small perturbation at the input layer will result in a totally wrong classification, which is so-called adversarial examples\cite{gu2014towards,kurakin2016adversarial,szegedy2013intriguing}.

Adversarial examples bring forth potential security threats for AI applications. Depending on where the adversarial examples are generated, there are the following two main situations.  One situation assumes a threat model, in which adversarial examples can be directly imported into the machine learning models. There has been lots of prior work proved that the adversarial examples can be generated by adding fine-grained modification to the original data in areas such as sound, image and text\cite{carlini2016hidden,dalvi2004adversarial,szegedy2013intriguing}. The other situation is physically realizable attacks, where adversarial examples are input into the AI applications that operate in the physical world. Up to now, there also has been many researchers tried to generate adversarial examples in the fields of image classification and face recognition\cite{kurakin2016adversarial,sharif2016accessorize}.

However, few studies have focused on the adversarial examples in the domain of robot path planning. With the rise of deep learning models, Deep Reinforcement Learning (DRL) algorithms achieve a state-of-the-art performance in the domain of robot path planning, such as VIN, DQN and A3C\cite{panov2018grid,tamar2016value,zhu2017target}. The following question thus arises: is there similar adversarial examples that may attack the DRL algorithms in the domain of robot path planning?

\begin{figure}
  \centering
  \subfigure[the normal map]{
    \label{fig:maps:a} 
    \includegraphics[width=4cm]{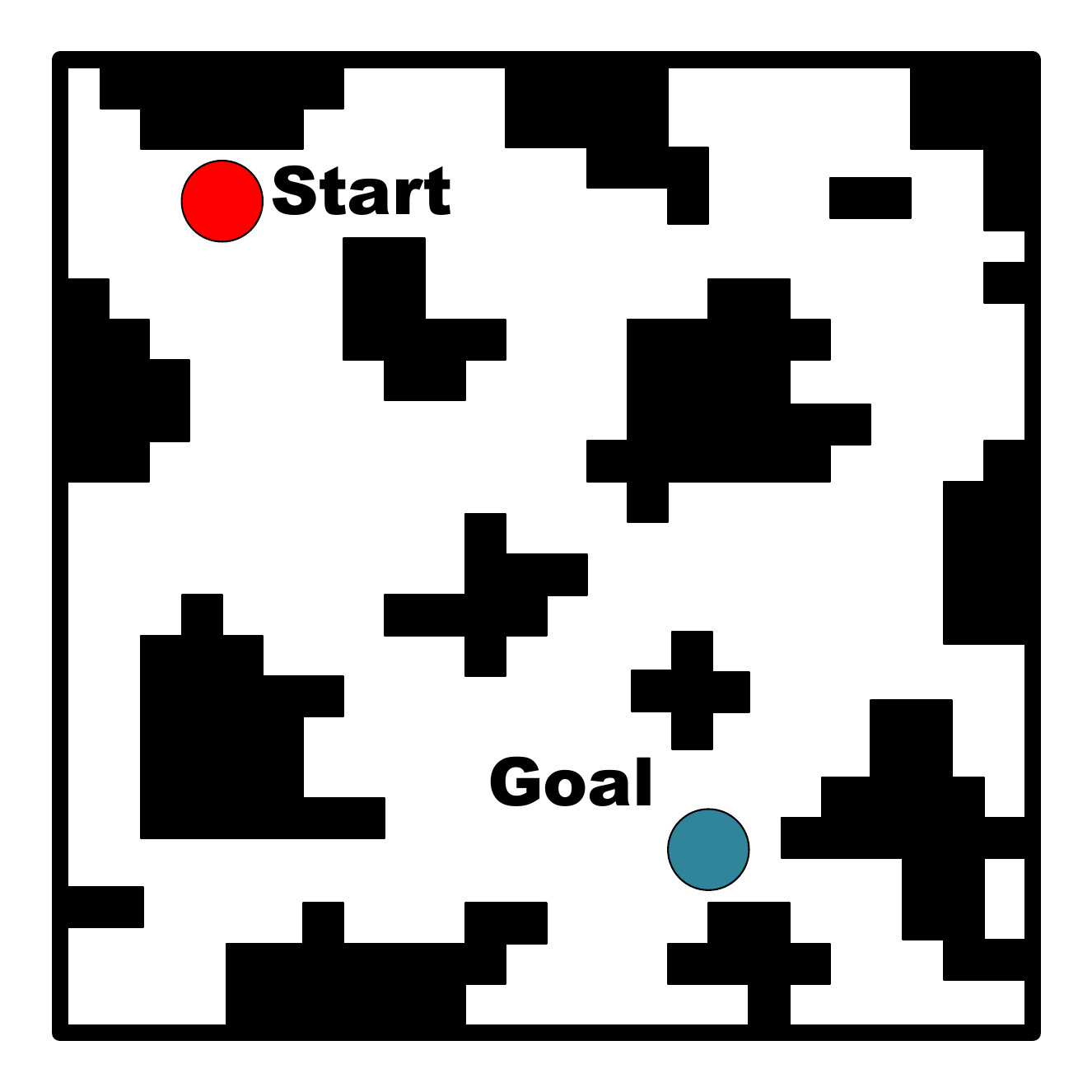}}
  \subfigure[the adversarial map]{
    \label{fig:maps:b} 
    \includegraphics[width=4cm]{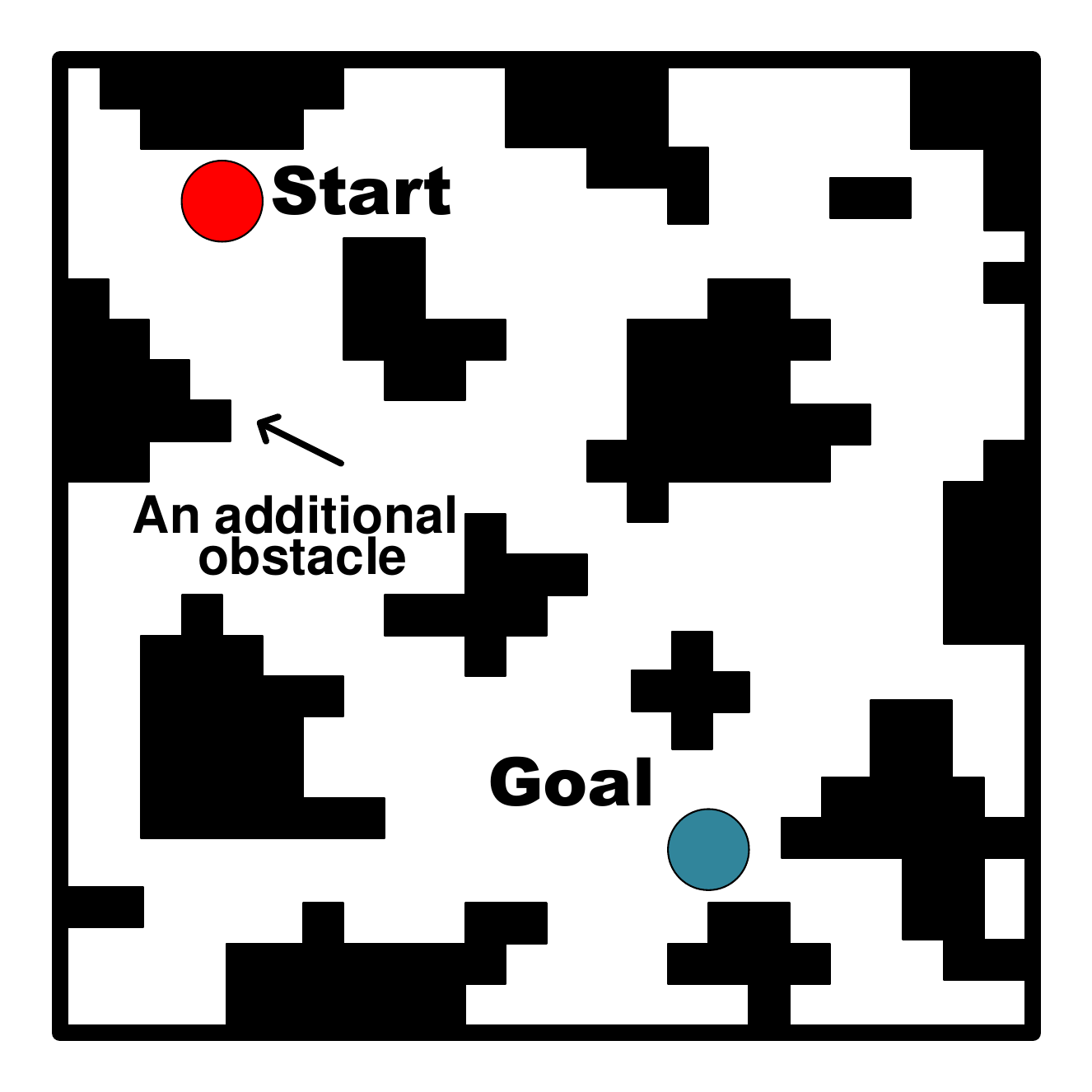}}
  \caption{A pair of maps:(a) is the normal map with random obstacle configurations; (b) is the adversarial map generated by attackers, which is the same as (a) except for an additional obstacle.}
  \label{fig:maps} 
\end{figure}

Some prior work has found the adversarial examples that exist in the process of robot path planning using some DRL algorithms, such as DQN, A3C, and VIN~\cite{bai2018adversarial,chen2018gradient,liu2017method}. From the literature review, we can find some commonalities between the adversarial examples. In the process of robot path planning, the algorithm can obtain the global map as the input data, and the task is to move from the start to the goal avoiding the obstacles. Therefore, an adversarial example for the path planning domain would consist of obstacle-level points adding to the target map (see Fig.~\ref{fig:maps}), so that a human observer would recognize it as a normal map, but a machine learning algorithm would plan a very different path for it.

In this paper, we specially focus on the adversarial examples of VIN path planning, which can achieve a better  generalization compared with other DRL algorithms~\cite{tamar2016value}. In the paper~\cite{liu2017method}, they have proposed a method that can generate adversarial maps for VIN path planning by adding obstacles to the grid-world domain. This method can automatically add obstacles based on the three rules they defined to obstruct the process of VIN path planning.

However, there are some deficiencies in the adversarial examples identification in~\cite{liu2017method}. Since it is not guaranteed that the generated adversarial maps will obstruct the VIN path planning successfully, they evaluate the attack results through a manual observation method. More specifically, they manually compared the two VIN-predicted paths to observe whether there are differences between them. This manual evaluation method is time-consuming and ineffective. For example, when manually observing the path on each map, one need to compare the paths with eyes, which can take at least a few seconds for each pair of maps. Especially when generating a large amount of adversarial maps, it's not possible to quickly get feedback about the effect of attacks through manual observation. Therefore, it is necessary to explore a method to identify the VIN adversarial examples automatically.

Aiming at the existing problem of the manual observation method, we explore a fast approach to automatically identify VIN adversarial examples. As previously mentioned, in order to estimate whether an attack is successful, we need to compare the difference between the two paths on a pair of maps. Therefore, we analyze the possible scenarios of the adversarial maps and define the categories of the predicted pairs. In order to realize a large-scale detection, we extract the two paths into one path image, and transform the path comparison to path image classification according to defined categories. But the classifier may be confused by the various shapes of the paths, so we also retain those easily distinguishable path feature to help the classifier to achieve a high accuracy. In this way, we implement a fast and high-accuracy identification method for VIN adversarial examples by combing the path feature and path images.

The rest of the paper is structured as follows: In Section 2, we discuss the related works in this field. This is followed in Section 3 by details about the categories definition of adversarial maps. Section 4 describes the method we propose for identifying the VIN adversarial examples automatically. Finally, Section 5 describes our experiments with the adversarial maps dataset.

\section{Related Work}
Since this paper is mainly focus on the adversarial examples in the domain of robot path planning, aiming to achieve an automatic detection of VIN adversarial examples. Therefore, in the related work of this paper, we investigate the existing research work in the field of adversarial examples, and discuss the feasible methods to identify adversarial examples in the domain of path planning.

 \subsection{Adversarial Examples in Applications}
From the literature review, the research of adversarial examples mainly focus on the following three fields of AI applications.
\paragraph{\textbf{In image classification}} Most of prior work pay attention to the adversarial examples in the field of image classification, which are generated by adding fine-grained per-pixel modifications to the input images~\cite{goodfellow2018explaining,gu2014towards,kurakin2016adversarial,szegedy2013intriguing}. As a result, the adversarial examples generally will lead to a mis-classification.

\paragraph{\textbf{In speech recognition}} There also has been researchers generating hidden voice commands to attack the speech recognition system using synthesized obfuscated speech~\cite{carlini2016hidden}. In the field of speech recognition, the adversarial examples will cause a wrong voice command executed by the device.

\paragraph{\textbf{In Atari game}} The Atari games using DRL algorithms may also be attacked by the adversarial examples, which are generated by adding noise to the background of the game~\cite{huang2017adversarial}. As a result, the adversarial examples will make the Atari game not work.

In summary, the adversarial examples cause different attack results in different application fields. For some applications, the attack results are easy to estimate, such as image classification, where there are just correct classification or not. However, in the field of robot path planning, the adversarial maps cause various impact on the path. Therefore, we need to explore a method to compare the difference between the two paths to automatically estimate the attack results.

\subsection{Adversarial Examples identification in Path Planning}
In order to compare the difference between the two paths, we do some research about the related methods for the path comparison.

\paragraph{\textbf{Trajectory Similarity Based Methods} } There are already many mature algorithms to calculate the trajectory similarity, such as Hausdorff distance, Frechet distance and DTW algorithms\cite{toohey2015trajectory}. Or there are some ways to transform the problem, such as calculating the area enclosed by the two paths, or calculating the longest common subsequence(LCSS)\cite{magdy2015review}. Generally, these algorithms need to traverse the whole coordinate points to calculate the trajectory similarity, which require a large amount of calculation and time.
\paragraph{\textbf{Path Image Based Methods} } A representative algorithm for combining path feature and images classification is proposed by \cite{graham2013sparse}, which enhances the images with signature information to recognize handwriting characters. This is followed by a algorithm for Chinese character identification\cite{yang2015chinese}, which recognize the Chinese characters as finite paths. These methods avoid complex trajectory information but use the path feature to enhance the machine learning to automatically identify the images.

Inspired by these methods, this paper combines the path feature comparison and path images classification. We visualize the two VIN-predicted paths of a pair of map into one image, transforming the different cases of attack results into different categories of path images. At the same time, in order to avoid the situation where the path image classifier is confused by the various shapes of paths, we also do path feature comparison. In this way, we can achieve a high-volume and high-accuracy identification for VIN adversarial examples.

\section{Preliminaries}
 In this section, we analyse the possible impact of adversarial maps and define the four categories of adversarial maps. In this way, we can transform the different attack results to path images categories.
\subsection{Definition of Path Pairs}
VIN can predict a path for robot from start to goal position avoiding the obstacles, according to the map with obstacle configurations. In this paper, the map is a synthetic $28\times 28$ grid-world domain with randomly placed obstacles. Each step in the path can advance in 8-directions.

\begin{figure}[!htbp]
  \centering
  \subfigure[the original path]{
    \label{fig:paths:a} 
    \includegraphics[width=5.5cm]{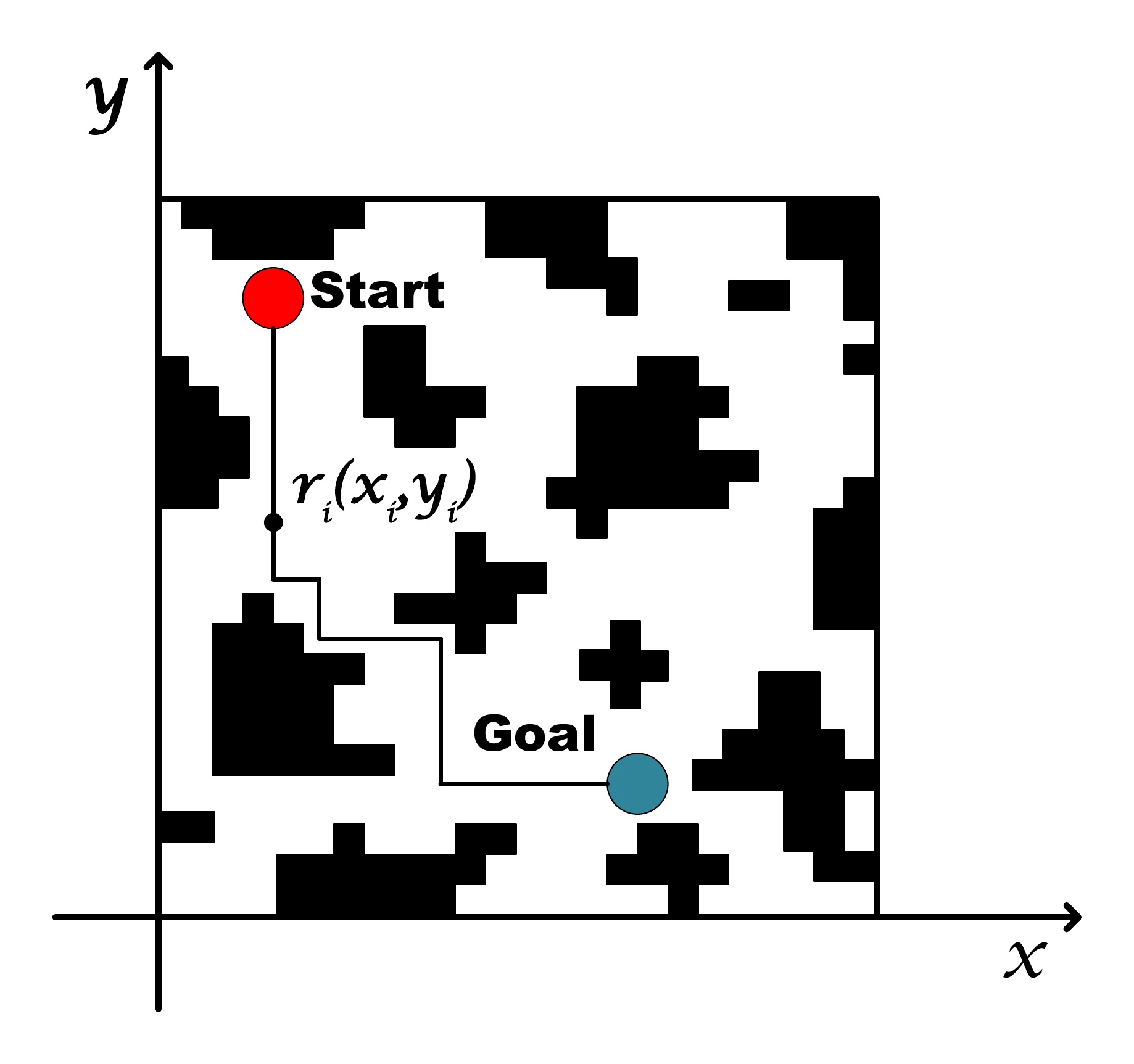}}
  \subfigure[the adversarial path]{
    \label{fig:path:b} 
    \includegraphics[width=5.5cm]{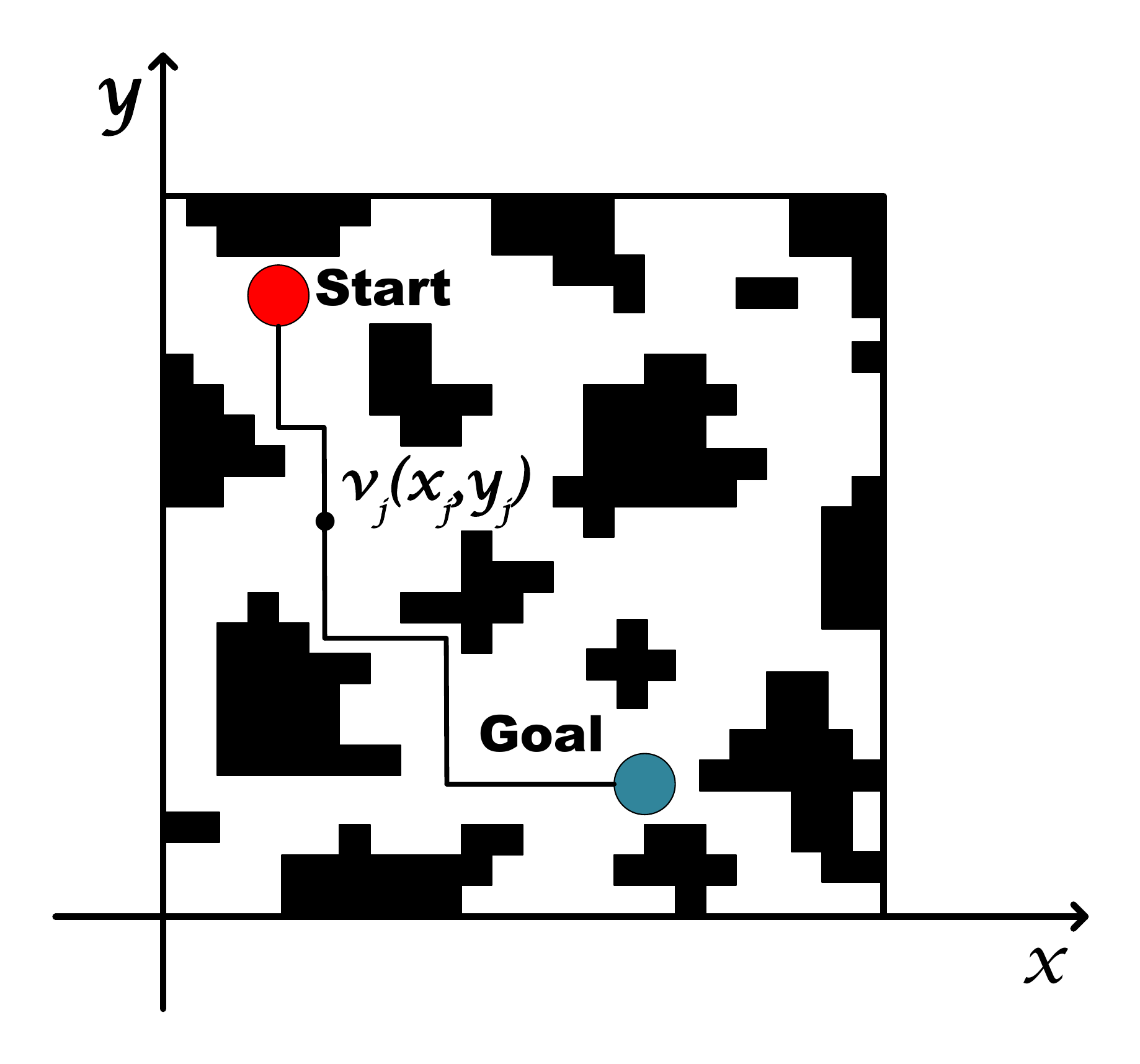}}
  \caption{A pair of paths: (a) is the original path which is predicted by VIN for the normal map; (b) is the adversarial path for the adversarial map. }
  \label{fig:paths} 
\end{figure}

We refer to an original map and an adversarial map as a pair of maps. In this way, we recognize the VIN-predicted trajectories for a pair of maps as a pair of path, the original path and the adversarial path. We describe the pair of path in a Cartesian coordinate system, as shown in Fig.~\ref{fig:paths}.

The paths can be expressed as follows:
\begin{equation}\label{1}
P_{original}=(r_1,r_2,r_3,...,r_n)
\end{equation}
where $P_{original}$ represents the VIN-predicted path for the original map, $r_i$ is the $i$ th step of the path, which can be expressed as $(x_i,y_i)$ in a Cartesian coordinate system.
\begin{equation}\label{2}
P_{adversarial}=(v_1,v_2,v_3,...,v_m)
\end{equation}
where $P_{adversarial}$ represents the VIN-predicted path for the adversarial map, $v_j$ is the $j$ th step of the path, which can be expressed as $(x_j,y_j)$ in a Cartesian coordinate system. And $n$ and $m$ are not necessarily equal.

Therefore, the definition of a pair of path is as follows:
\begin{equation}\label{3}
PathPair=(P_{original},P_{adversarial})
\end{equation}

Below we use the relationship between these pairs of paths to define the four categories of adversarial maps.

\subsection{Categories Definition of VIN Adversarial Examples}
\begin{figure}[!htbp]
  \centering
  \subfigure[the UrP]{
    \label{fig:cases:a} 
    \includegraphics[width=2.6cm]{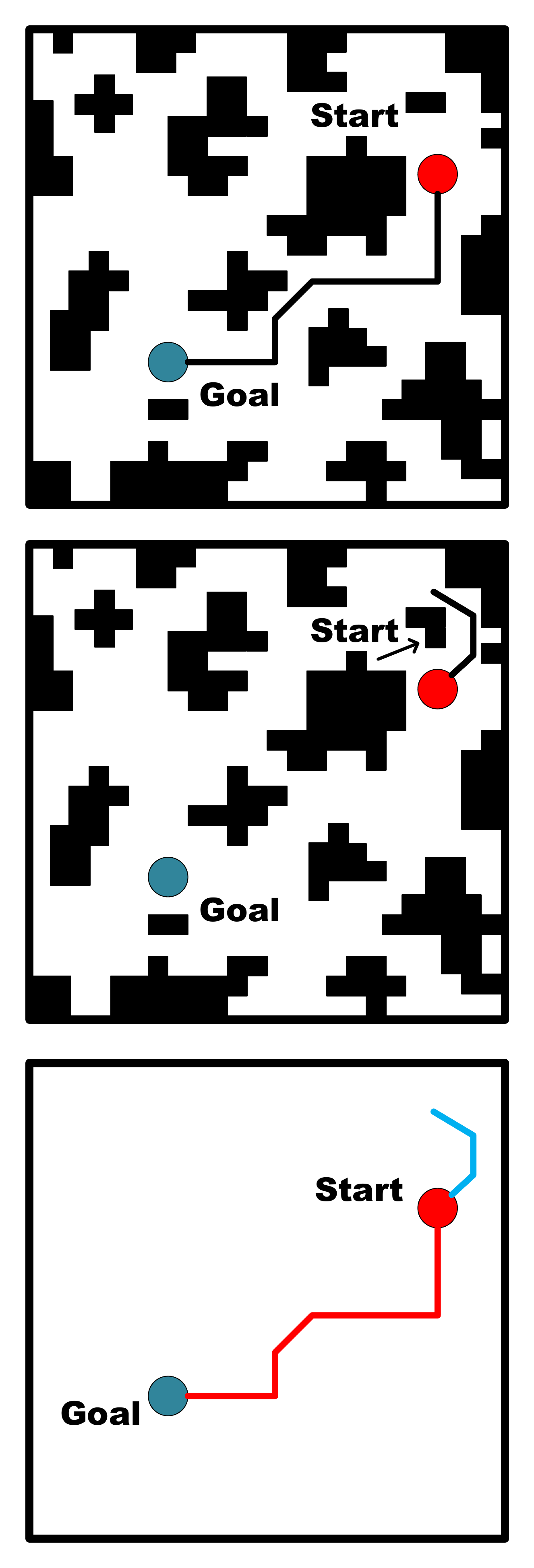}}
  \subfigure[the FP]{
    \label{fig:cases:b} 
    \includegraphics[width=2.6cm]{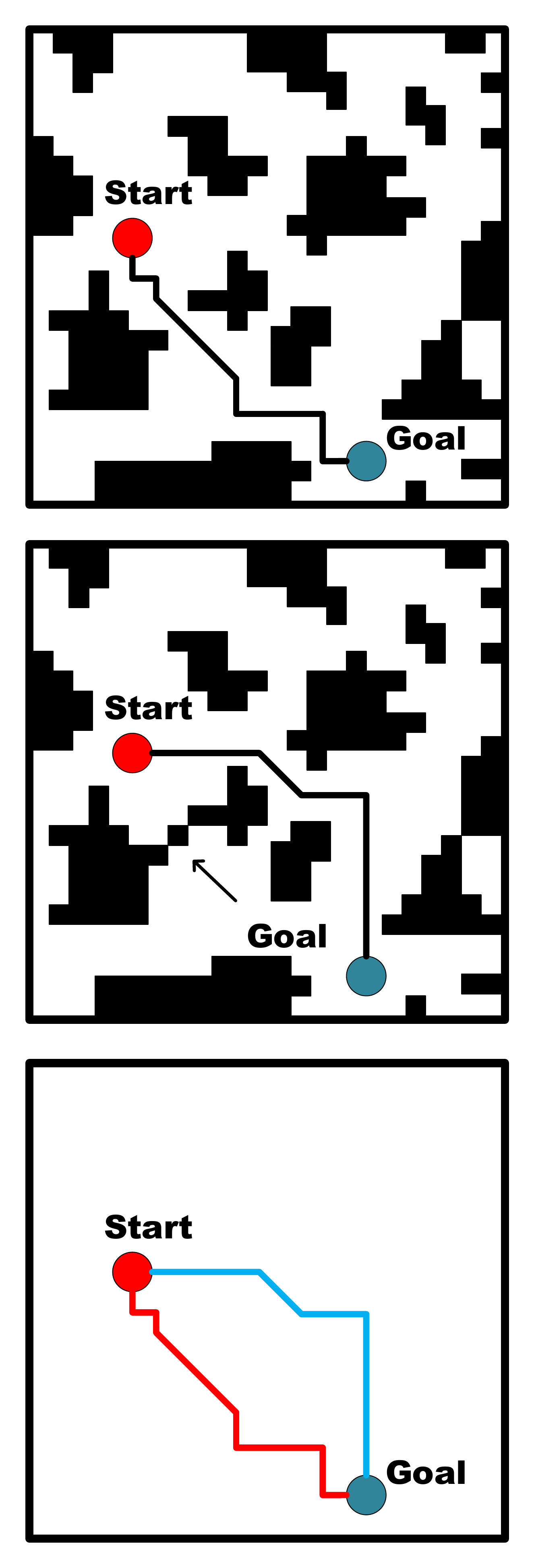}}
    \subfigure[the DP]{
    \label{fig:cases:c} 
    \includegraphics[width=2.6cm]{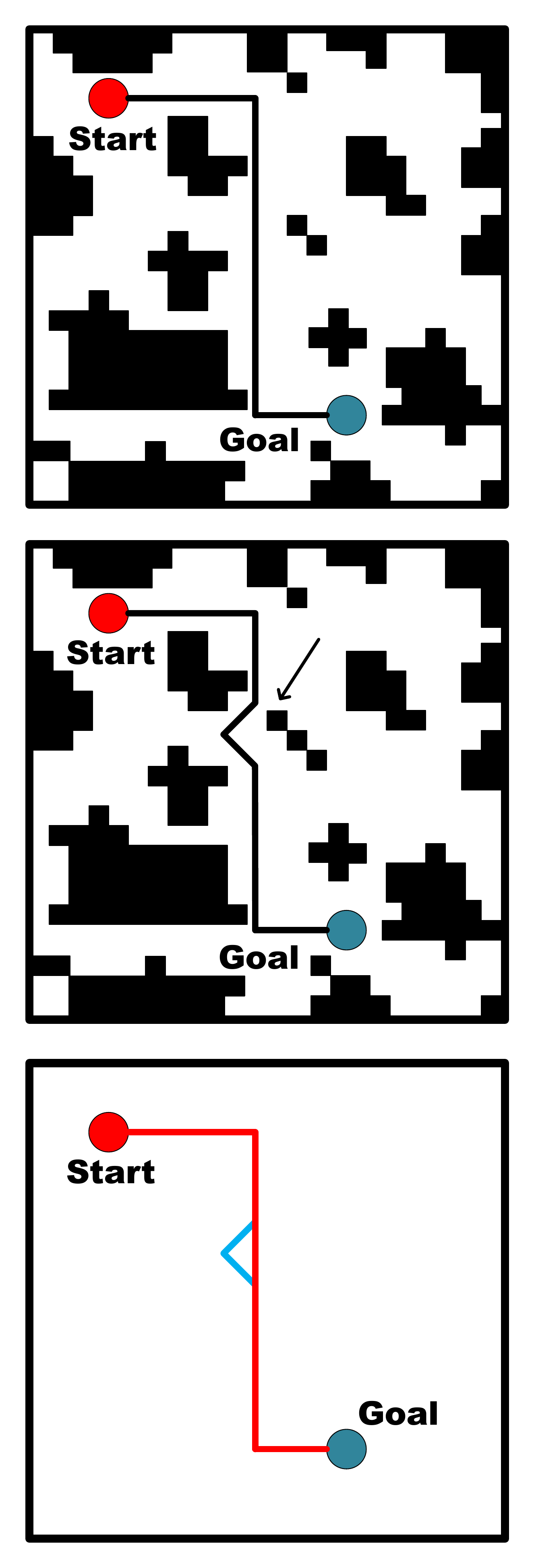}}
      \subfigure[the UcP]{
    \label{fig:cases:d} 
    \includegraphics[width=2.6cm]{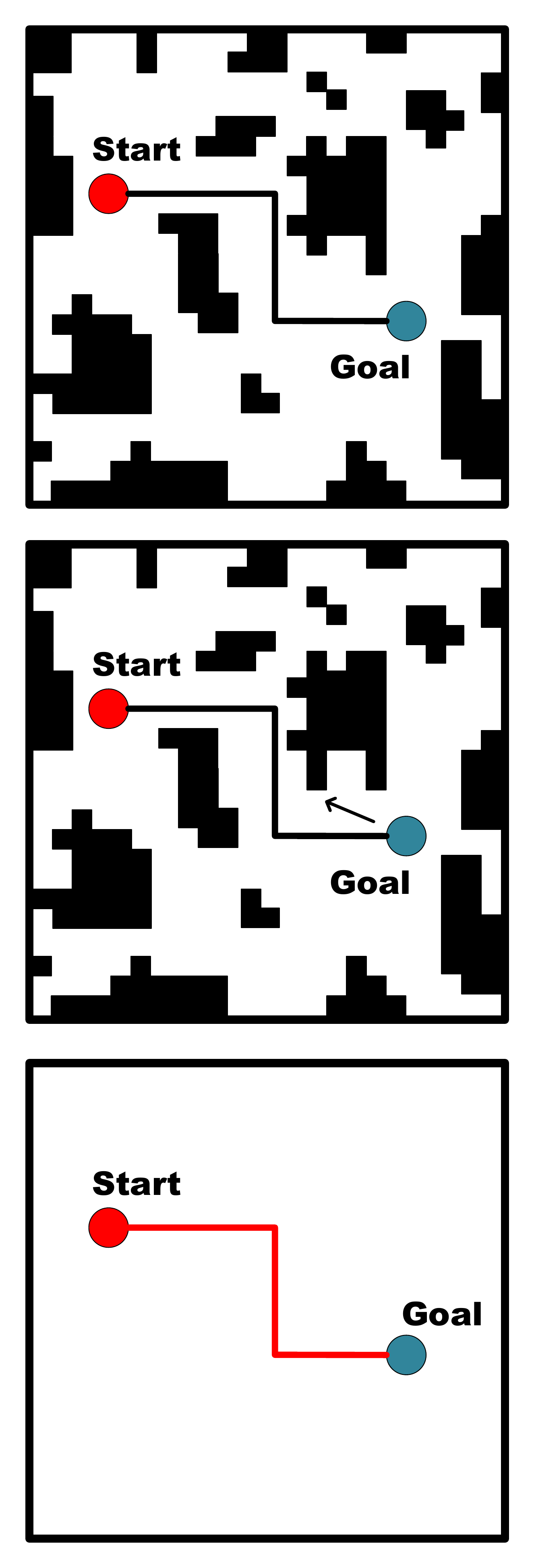}}
  \caption{Four categories of VIN adversarial maps. The first line is the original maps, the second line is the adversarial maps, and the third line is the extracted path image.}
  \label{fig:cases} 
\end{figure}

As it came to be known, the adversarial maps may cause various impact on the predict path. By visualizing the pair of paths on a path image, we transform the different attack results into different categories of path images. We divide the results into four limited categories as shown in the Fig.~\ref{fig:cases}

From the visual observation, it is obvious to find the differences of the four categories. In the unreached path (UrP) class, the adversarial paths can't reach to the goal position. In the fork path (FP) class, the two paths are very different, forming a fork in the path. In the detour path (DP) class, the adversarial maps do not affect the general direction of the path, where the adversarial paths just simply bypass the additional obstacles. In the unchanged path (UcP) class, the two paths are exactly the same, which means the adversarial maps do not impact VIN path planning. We give the definition of these four cases based on the description of the path pairs.

\paragraph{\textbf{Four Categories of Adversarial Maps}}
The impact of adversarial maps on VIN path planning can be divided into four limited categories based on  the difference between $P_{original}$ and $P_{adversarial}$, which are UrP, FP, DP, and UcP.

\textit{\textbf{The Unreached Path (UrP)}}
$P_{adversarial}$ didn't reach to the goal position. That is
\begin{equation}
r_n(x_n,y_n)\not=v_m(x_m,y_m)
\label{case1}
\end{equation}

\textit{\textbf{The Unreached Path (UrP)}}
There is a significant difference between $P_{original}$ and $P_{adversarial}$, where the maximum vertical distance and the maximum horizontal distance in the part where $P_{original}$ and $P_{adversarial}$ do not coincide are greater than four. That is
\begin{equation}
\begin{split}
 \max\left|x_i-x_j\right|>4 &\text{ where } y_i=y_j  \text{ AND }  \max\left|y_i-y_j\right|>4  \text{ where } x_i=x_j \\
&  \text{ for }  i,j \text{ of the part where } P_{original} \neq P_{adversarial}
\end{split}
\label{case3}
\end{equation}


\textit{\textbf{The Detour Path (DP)}}
There is a slight difference between $P_{original}$ and $P_{adversarial}$, where the maximum vertical distance and the maximum horizontal distance in the part where $P_{original}$ and $P_{adversarial}$ do not coincide are less than or equal to four. That is
\begin{equation}
\begin{split}
 \max\left|x_i-x_j\right|\lid4 &\text{ where } y_i=y_j  \text{ AND }  \max\left|y_i-y_j\right|\lid4  \text{ where } x_i=x_j \\
&  \text{ for }  i,j \text{ of the part where } P_{original} \neq P_{adversarial}
\end{split}
\label{case3}
\end{equation}


\textit{\textbf{The Unchanged Path (UcP)}}
$P_{adversarial}$ is exactly the same with $P_{original}$. That is
\begin{equation}
P_{original} =P_{adversarial}
\label{case4}
\end{equation}

It should be additionally noted that the critical value 4 is not an constant, but is determined by the size of the map. When the domain becomes larger, the adversarial maps may cause a larger fork in the path.

In this paper, it is considered that an adversarial map which can make a significant change in the original path is a successful attack. Therefore, the adversarial maps that cause the UrP and FP results are the VIN adversarial examples that attack successfully. Correspondingly, the adversarial maps that cause the DP and UcP results fail to attack the VIN path planning.


\section{Training-based identification for VIN adversarial examples}
In this section we will introduce the identification approach to VIN adversarial examples. As mentioned previously, unlike other areas, the adversarial examples in the domain of path planning will produce multiple attack results on the original path. Therefore we identify the VIN adversarial examples according to the pathpairs classification results.

\begin{figure}[!htbp]
  \centering
  \includegraphics[width=12.3cm]{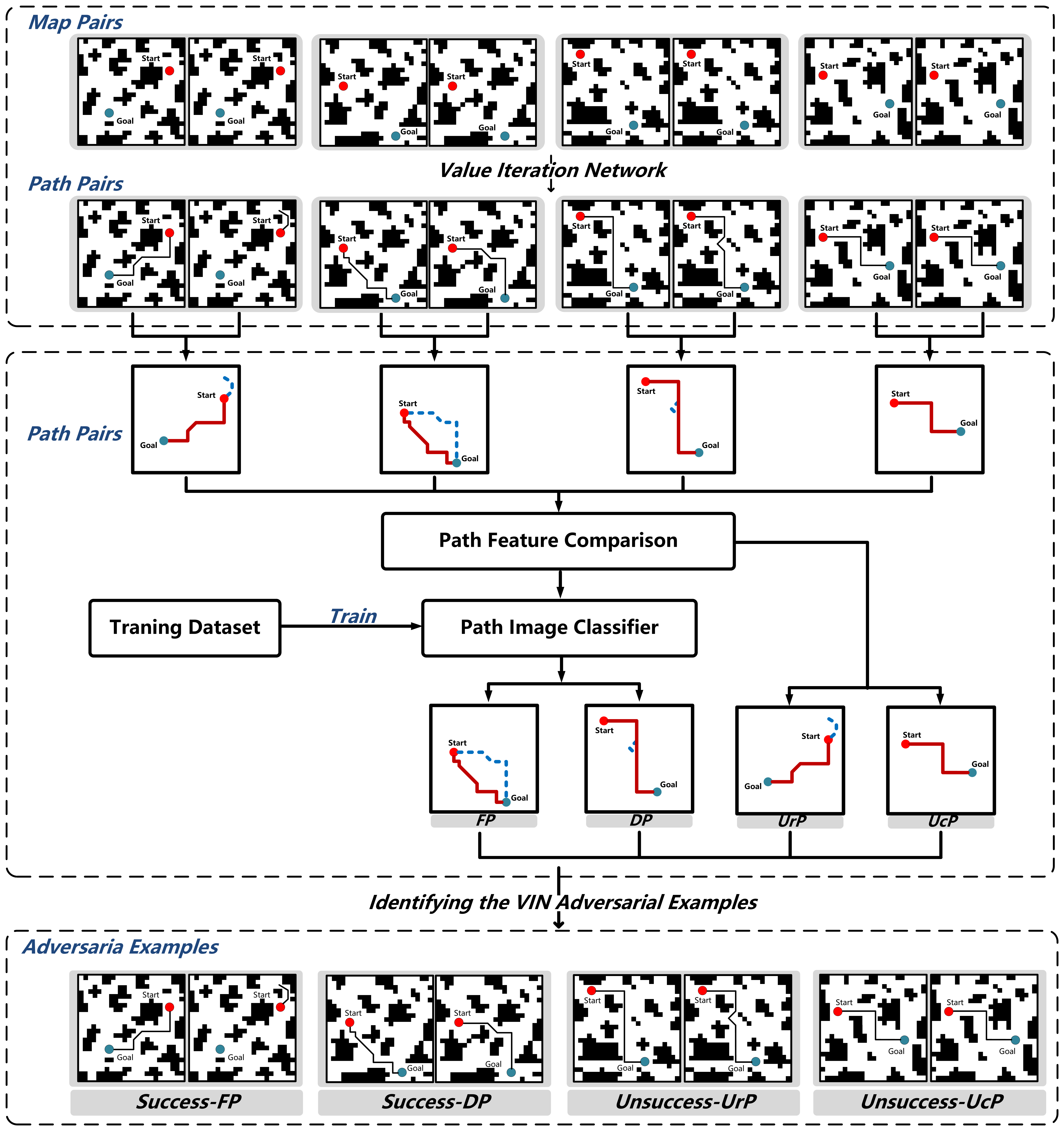}
  \caption{The framework of VIN adversarial examples identification approach}
  \label{fig:framework} 
\end{figure}
\subsection{Framework}
The framework of the method is mainly divided into three modules, as shown in the Fig.~\ref{fig:framework}. The first module is to perform VIN algorithm on the map pairs to generate the path pairs, which is called \textsl{the VIN path planning module}. Then, the second module is to classify the four categories of the adversarial maps by combining path feature comparison and path image classification, which is called \textsl{the classification module}. At last, the classification results of the path images are matched with the generated map pairs to identify the VIN adversarial examples, which is called \textsl{the identifying module}. Below we will introduce these three modules.

\paragraph{\textbf{VIN Path Planning Module}}
The generated map pairs are not guaranteed to attack the VIN path planning successfully. A map pair contains an original map and an adversarial map. The information on the original map, such as obstacle configurations, start point and end point, are all random. And the adversarial map is generated by adding an additional obstacle to the original map. After VIN path finding module, there will be a path pair corresponding with a map pair predicted by VIN. By studying the results of VIN path planning, we can know whether the adversarial maps have attacked successfully.
\paragraph{\textbf{PathPairs Classification Module}}
Classifying the path pairs is the most important module of our method. It is mainly divided into two parts, the comparison of path features and the classification of path images. Firstly, we classify the UrP and UcP categories by comparing the obvious path features. For the remaining path pairs where path features are not obvious, we transform them into path images. Then we classify the FP and DP categories by training a path image classifier. In this way, we classify the four categories and automatically identify them as FP, DP, UrP and UcP.
\paragraph{\textbf{Identification Module}}
In order to achieve an automatic detection of VIN adversarial examples, the final step in the method is to identify the adversarial examples through the labels of the path pairs. We need to match the label of the path pairs with the corresponding map pairs to determine whether the adversarial maps successfully attack VIN, and then identify the VIN adversarial examples. Therefore, in the identifying module, we finally identify the map pairs as Success-FP, Success-UrP, Unsuccess-DP and Unsuccess-UcP.

\subsection{Training-based Classification for Path Pairs}
The main body of the algorithm is the classification module for the path pairs. As described in Section 3, it is easily to find that UrP and UcP can be well distinguished by path features, but it will take a lot of time to compare the path features for DP and FP. Whereas, DP and FP can be well recognized by the path images. The two paths of FP are very different, presenting an irregular polygon shape, while the two paths of DP are similar, presenting a line shape. But UrP and UcP are difficult to be recognized by the path images, for the two pathes generally present a line shape. Therefore we combine of the path feature and the path image to do the classification. We identify the UrP and UcP through path feature comparison and identify the DP and FP through path image classification. This allows for a large-scale and high-accuracy classification.

\begin{figure}
  \centering
  \includegraphics[width=10cm]{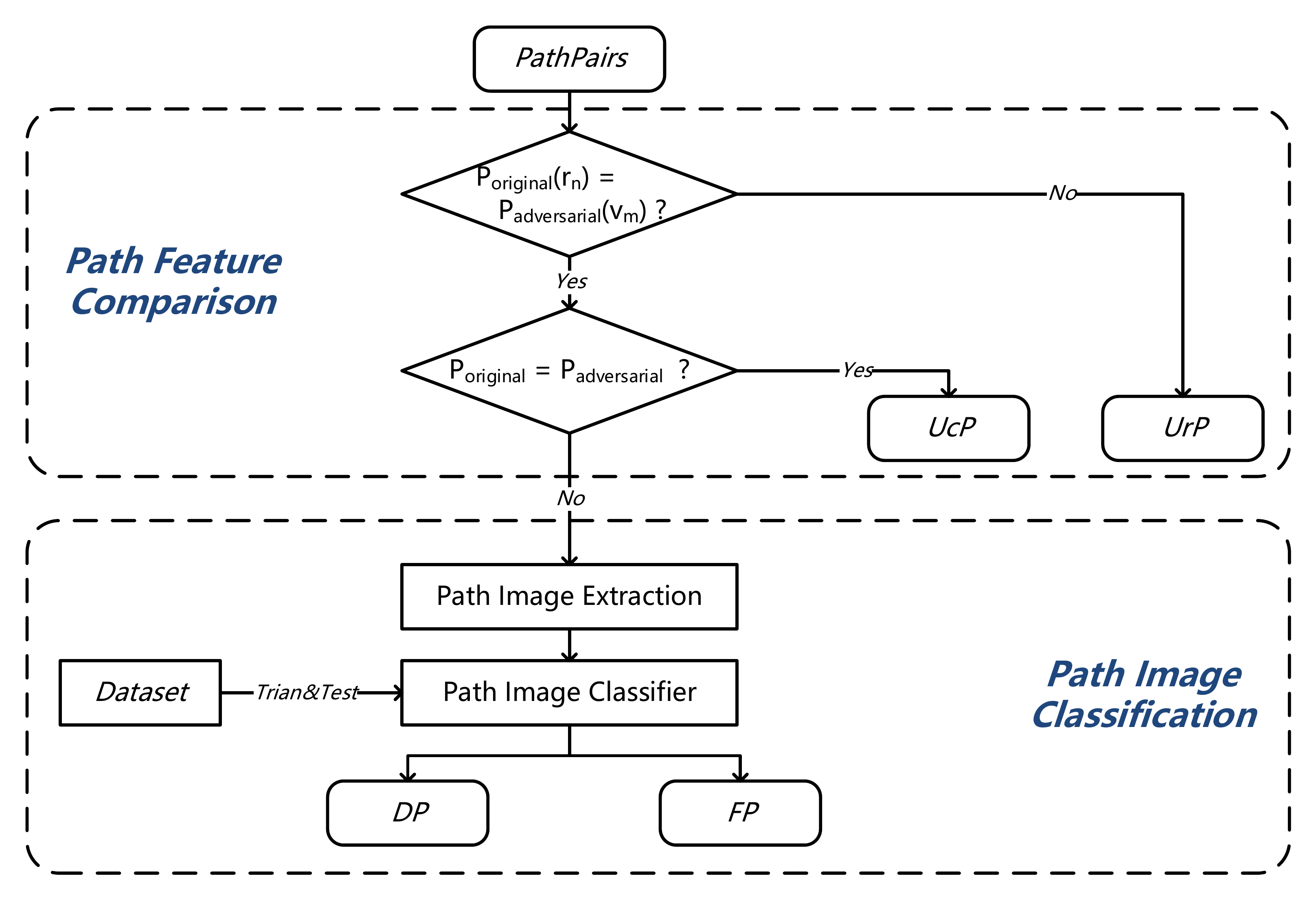}
  \caption{The flow chart of classification module.}
  \label{fig:comparison} 
\end{figure}

\paragraph{\textbf{Path Feature Comparison}}
We obtain the path pair of the map pair after the VIN path planning module, which contain all coordinate points from the start to the goal position. Firstly, we compare whether the last point in the two paths is equal. If they are not equal, which means the adversarial path fails to reach the goal, the path pair belongs to UrP category. Then we compare whether the points in two paths are completely equal. If they are equal, which means the path has not been changed, the path pair belongs to UcP category. In this way, through the obvious features of the paths, the UrP and UcP classes can be quickly distinguished. Path pairs that cannot be classified by these two features belong to the remaining DP and FP categories.

\paragraph{\textbf{Path Image Classification}}
For the DP and FP categories, the difference between the paths is not obvious, and the coordinate points in the path cannot be one-to-one correspondence. However, it can be seen from the path images that the difference is obvious in the path shape. So we use the path image to classify the DP and FP categories. We build a path image classifier, which is trained and tested using the labeled path image database. And we use the built classifier to classify the remaining path images of DP and FP categories.
\subsection{Path Image Classifier Building}
To build a path image classifier, we need to train the classifier with the path image dataset as shown in Fig.~\ref{fig:comparison}. After the path map extraction process, the path pairs are transformed into path images. We use the labeled path image dataset to train several classifiers in order to select the classifier that best fits the path image data. The built path map classifier will be used to classify new path image data.

\paragraph{\textbf{Path Image Extraction}}
In the path map extraction process, we visualize each pair of path in one path image. The pair of path are visualized with different colours, where path in red represents the original path and blue represents the adversarial path. In the portion where the two paths coincide, the red line covers the blue line. It means that the blue path in a path image represents the part of the adversarial path different from the original path. We visualize the path pair on the same $28\times 28$ blank map to prevent obstacles affecting the classification. Therefore, after the process of path image extraction, the same size path images are generated for classifiers.

\paragraph{\textbf{Classifier Selection}}
For the selection of path image classifier, we mainly consider two categories of classifiers, traditional classifiers and deep neural network classifiers. In recent years, deep neural networka have performed well in the field of image classification, especially Convolutional Neural Network (CNN)~\cite{russakovsky2015imagenet}. However, the training of CNN generally requires a large amount of dataset, which is difficult to obtain for VIN path planning. Therefore, we also consider the traditional machine algorithm that requires less training data, such as Support Vector Machine(SVM)~\cite{liu2017svm}. Unlike the usual images, there is not much pixel information in the path images. Nor can it be equivalent to character images, for the path shapes of each path image are different. So we compare several classifiers through experiments to select the most suitable classifier for path images.

\subsection{VIN Adversarial Examples Identification Method}
The method for VIN adversarial examples can be summarized as follow. In the algorithm, the input is the original map pairs $D_{map}$, and the output is the adversarial examples with label information $D_{adversarial}$. The three processes of the method are VIN path planning, path pair classification, and adversarial examples identification. We implement an automatic identification approach for VIN adversarial examples. The specific process steps of the algorithm are shown as Algorithm~\ref{alg:method}.

\begin{algorithm}[!htbp]
\caption{VIN Adversarial Examples Identification Method}
\begin{algorithmic}[1]
\REQUIRE $D_{map}=\{MapPair_1, MapPair_2, ...,MapPair_i, ... , MapPair_n\}$
\ENSURE {$D_{adversarial}=\{(MapPair,PathPair,Label)_1,...,(MapPair,PathPair,Label)_m\}$}
\STATE{$D_{map} \overset{VIN} \longrightarrow D_{path}={\{(MapPair,PathPair)\}}_{1}^{n}$}
\STATE {extract $D'_{path}={\{(PathPair)\}}_{1}^{n}$ from $D_{path}$}
\STATE {$D_{label}=\{(PathPair,Label)_1,(PathPair,Label)_2,...,(PathPair,Label)_n\}$}
\FOR{each $PathPair_i$ in $D_{path}$}
\IF{$P_{original}(r_n)\neq P_{adversarial}(v_m)$}
\STATE{$Label_i=UrP$}
\ELSIF{$P_{original} =P_{adversarial}$}
\STATE{$Label_i=UcP$}
\ELSE
\STATE{$D_{path\_image}=D_{path\_image}+PathPair_i$}
\ENDIF
\ENDFOR
\STATE {$D_{path\_image} \overset{Classifier} \longrightarrow$ $D_{label\_image}$}
\STATE {$D_{label}=D_{label}+D_{label\_image}$}
\STATE {Matching $D_{label}$ AND $D_{path} \longrightarrow D_{label}={\{(MapPair,PathPair,Label)\}}_{1}^{n}$}
\FOR{each $Label_i$ in $D_{label}$}
\IF{$Label_i=UrP$ OR $FP$}
\STATE{$D_{advesarial}=D_{advesarial}+(MapPair_i,PathPair_i,Label_i)$}
\ENDIF
\ENDFOR
\RETURN{$D_{advesarial}$}
\end{algorithmic}
\label{alg:method}
\end{algorithm}

\section{Experiment And Evaluation}
In our experiments, we generate map pairs using the method proposed in~\cite{liu2017method} as experimental dataset. Then we use the algorithm designed in Section 4 to identify the VIN adversarial examples. At last, we evaluate the feasibility and dependability of our method.
\subsection{Experiment Setup}
\begin{figure}[!htbp]
  \centering
  \includegraphics[width=12.3cm]{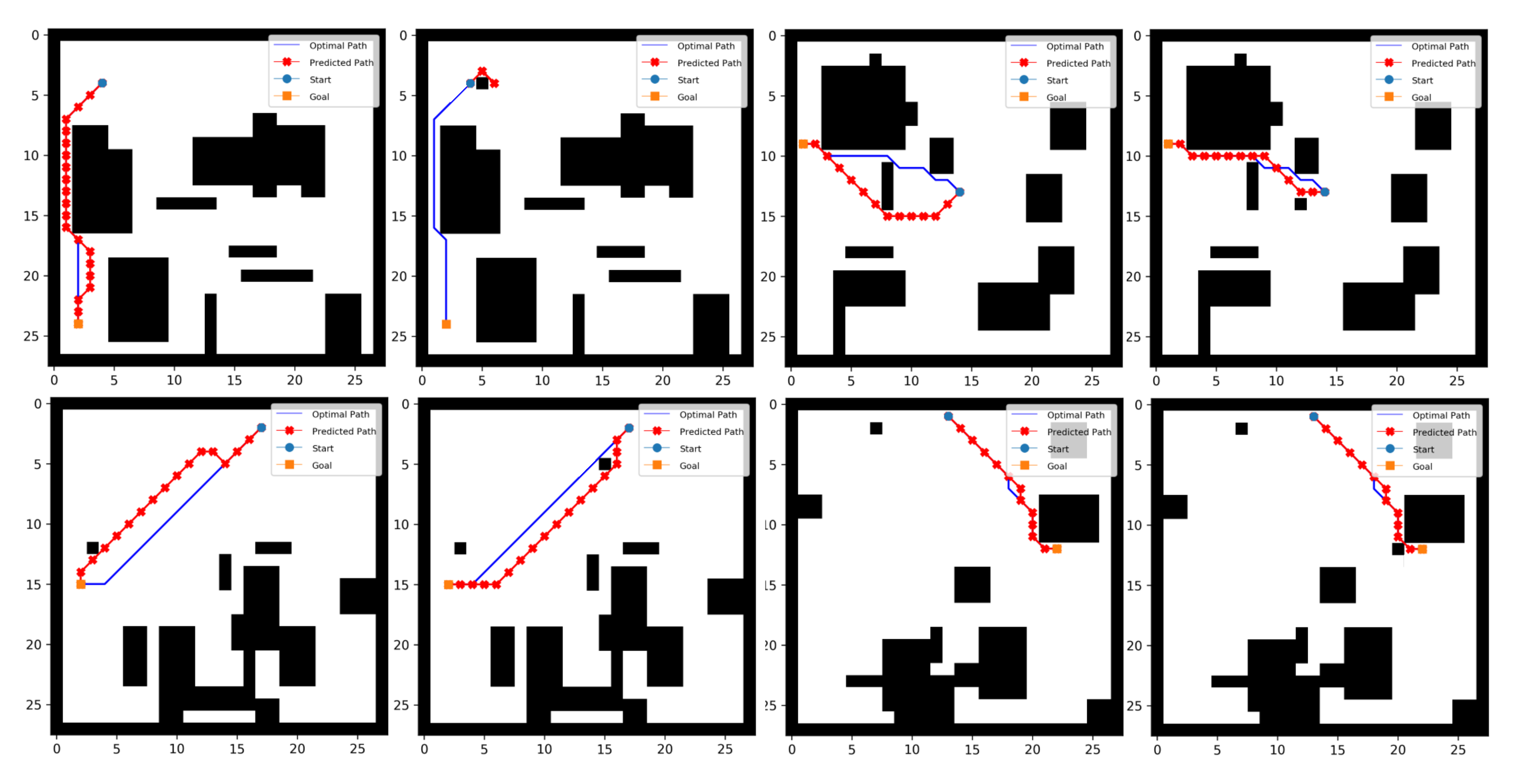}
  \caption{The examples of map pairs. }
  \label{fig:mappair} 
\end{figure}

The experiment setup mainly consists of the following steps:
\paragraph{\textbf{Experiment Environment}}
We perform our experiments on the environment shown as Table~\ref{tab1}, and using the source code provided by~\cite{tamar2016value} for VIN path planning.
\begin{table}
\centering
\caption{The experimental environment.}\label{tab1}
\begin{tabular}{r|l}
\hline\hline
  \textsl{ \bfseries No.    } & \textsl{  \bfseries  Environment  }\\
\hline
 { \bfseries Operation System } & { Windows 10 }\\
{\bfseries CPU } & {  Intel(R) Core(TM) i7-4510U @2.00GHz 2.60GHz  } \\
{\bfseries RAM } & { 8GB }  \\
{\bfseries Hard Disk }  & { 1T } \\
{ \bfseries  Programming Language }  &{ python 2.7 }\\
{ \bfseries  Database }  &{ mysql } \\
\hline\hline
\end{tabular}
\end{table}
\paragraph{\textbf{Data Preparation}}
We generate 5,000 pairs of maps. The original map is a $28\times 28$ grid-world domain, where the obstacle configurations and start and goal positions are all random. The adversarial map is generated by adding one additional obstacle to the original map. We use the VIN algorithm to predict paths for each pair of maps, so that we can obtain a pair of VIN-predicted path (see Fig.~\ref{fig:mappair}). Because it takes a long time for VIN to process a map, it is difficult to get a large number of path pairs. And since the map pair is generated randomly, 5,000 pairs of maps can already contain all the possible situations. So we used 5,000 map pairs as the dataset for our experiment.

\paragraph{\textbf{Data Storage}}
 In order to implement the matching of map pairs, path pairs and labels in the algorithm process, we store the original dataset in mysql database. And it is also convenient to study the adversarial examples in the future work. The structure of data table is designed as Table~\ref{tab2}. No is an incremented sequence number. The 0-1 matrix of the Map is drawn into a one-dimensional array, separated by symbols and stored as varchar type, which is the same for Path. Label contains the value \{UrP, UcP, DP, FP\}. Attack is a boolean value, where 1 means the attack is successful and 0 means the attack is unsuccessful.

 \begin{table}[!htbp]
\centering
\caption{The structure of data table}\label{tab2}
\begin{tabular}{c|ccccccc }
\hline\hline
\textsl{ \bfseries  Column Name } &{ \bfseries  No }  &{ \bfseries  Map\_O } & { \bfseries  Map\_A } & { \bfseries  Path\_O  } & { \bfseries  Path\_A  } & { \bfseries  Label } & { \bfseries  Attack } \\
\hline
\textsl{ \bfseries  Data Type } & Int &  Varchar & Varchar & Varchar & Varchar & Varchar & Boolean \\
\hline\hline
\end{tabular}
\end{table}

\begin{figure}[!htbp]
  \centering
  \includegraphics[width=11cm]{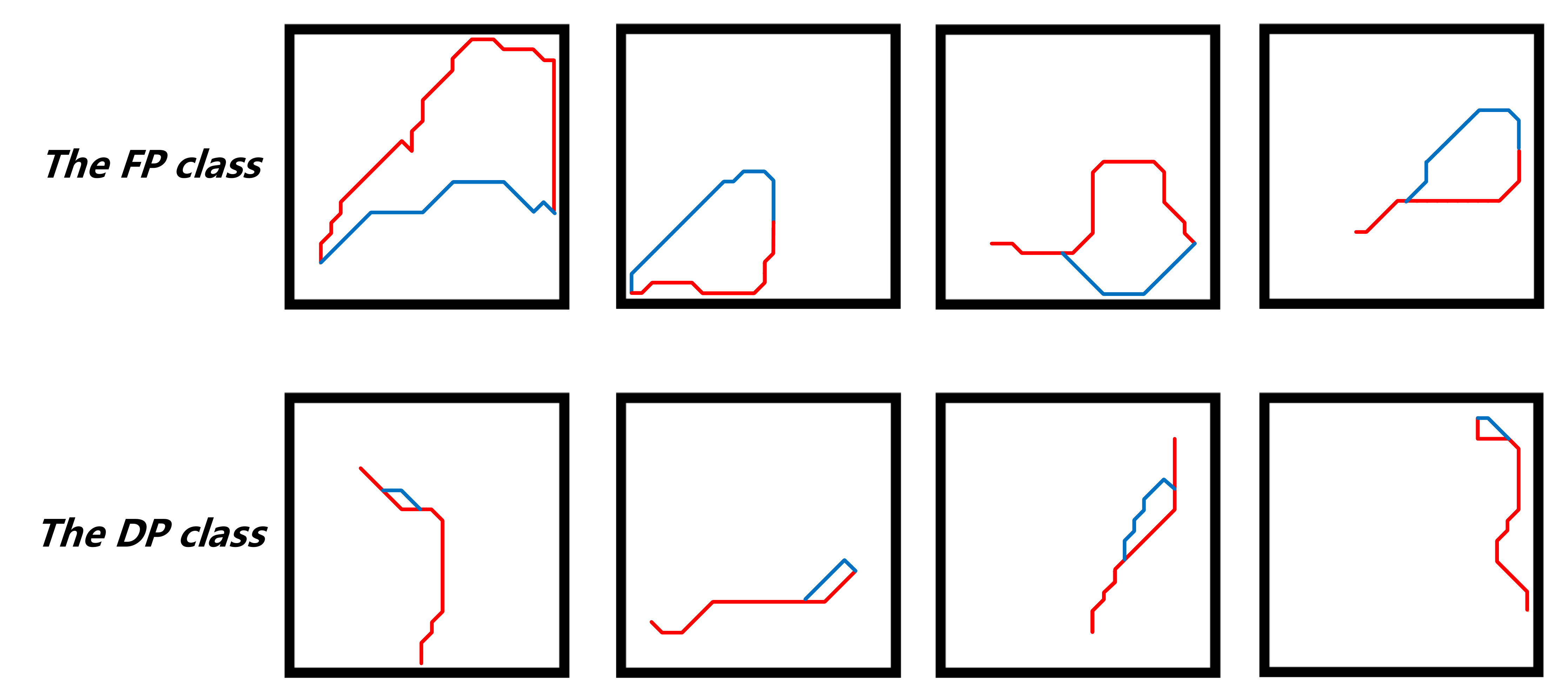}
  \caption{The path image database of DP and FP categories.}
  \label{fig:augmentation} 
\end{figure}

\paragraph{\textbf{Dataset Building}}
 In order to train the path image classifier, we need to build the path image dataset of DP and FP. We use the obtained 5,000 path pairs to generate 5,000 path images, and do image annotation for the DP and FP categories. The dataset of DP and FP is shown in the Fig.~\ref{fig:augmentation}.

\paragraph{\textbf{Experiment Evaluation}}
Since the classifier in the paper is a binary classification problem and the path image dataset is imbalance, we use the ROC curve and Precision-Recall curve to evaluate the confidence of the classifier accuracy. The larger the area under the curve, the higher the confidence of the classifier accuracy.

\subsection{Evaluation And Analysis}
We evaluate the dependability and feasibility of the method mainly from two aspects, classifier accuracy and processing time. We compare three commonly used classifiers to select the classifier that is most suitable for path image classification. We also compare the processing time of the training-based method and the manual observation method to evaluate the feasibility of our method.

The distribution of the 5,000 path pairs dataset is shown as Table~\ref{tab4}. It can be seen that most of them are UrP category that fail to attack VIN, and the data amount of FP category is the smallest.

\begin{table}[!htbp]
\centering
\caption{The data distribution of the four categories}\label{tab4}
\begin{tabular}{c|cccc}
\hline\hline
\textsl{  \bfseries Categories  }& \textsl{ \bfseries Unreached Path } &  \textsl{  \bfseries Fork Path  } &  \textsl{  \bfseries Detour Path }  &  \textsl{  \bfseries Unchanged Path }  \\
\hline
 \textsl{  \bfseries Quantity  } &  1,632 &  204 & 1,076 & 2,088 \\
\hline\hline
\end{tabular}
\end{table}
\subsubsection{Comparison of Classifiers}
In this paper, the dependability of the method mainly depends on the accuracy performance of the path image classifier. Therefore, we use the path image database to train and test several classifiers that commonly used to compare their performance.

\begin{figure}[!htbp]
  \centering
  \includegraphics[width=9cm]{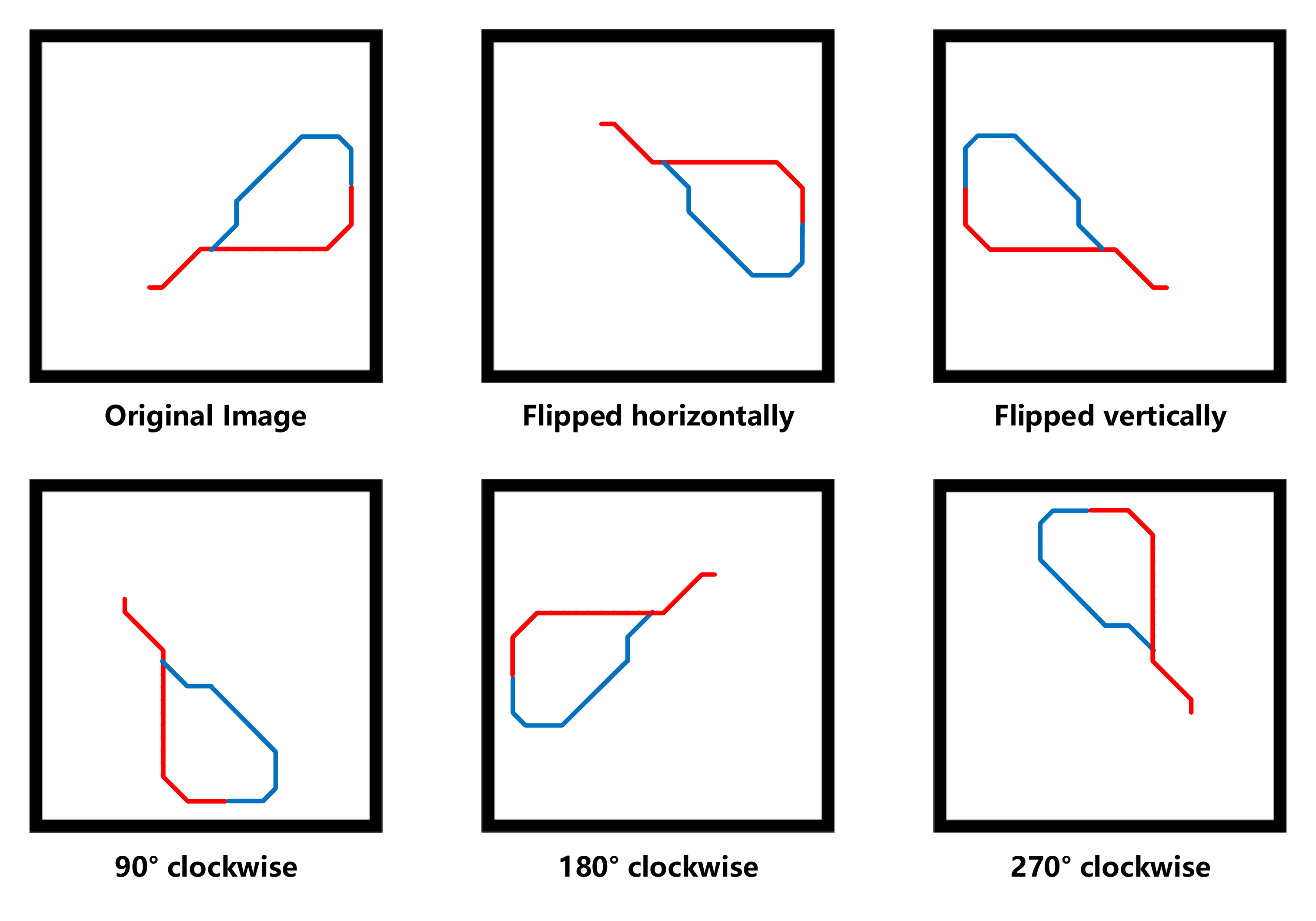}
  \caption{The augmentation for path images of FP category.}
  \label{fig:flip} 
\end{figure}

In order to avoid the impact of data imbalance on the classifiers, we firstly do the data augmentation for the path image database. As we can see from Table~\ref{tab4}, there is a data imbalance of DP and FP categories. It is because in the black-box environment, we can't control the attack results of adversarial maps. So we need to do data augmentation for the FP category to balance the path image dataset. In order not to affect the authenticity of the data as much as possible, we only flip the path images horizontally and vertically, and rotate images clockwise 90, 180 and 270 degrees  (see Fig.~\ref{fig:flip}). In this way, we can randomly generate the images for FP category to the same number as the DP category.

Then we train three classifiers based on our image path dataset, that are Support Vector Machine(SVM), Convolutional Neural Network(CNN) and Multilayer Perceptron(MLP). We compare their performance in the following four aspects.

\paragraph{\textbf{Comparison 1: parameters}}
The basic structure of both MLP and CNN is neural network. Taking CNN as an example, it is determined by multiple parameters such as filter size, number of convolution layers, pooling size, optimizer, learning rate and so on. Moreover, the relationship between parameters is complex. So it is difficult to determine which parameter has a greater influence on the classification. However, the parameters of the SVM are much fewer. To determine an SVM classifier, we only need to select a type of kernel and adjust the related parameters of kernel. Obviously, the parameters of SVM are easier to control and adjust than neural networks.

\begin{figure}[!htbp]
  \centering
  \includegraphics[width=8cm]{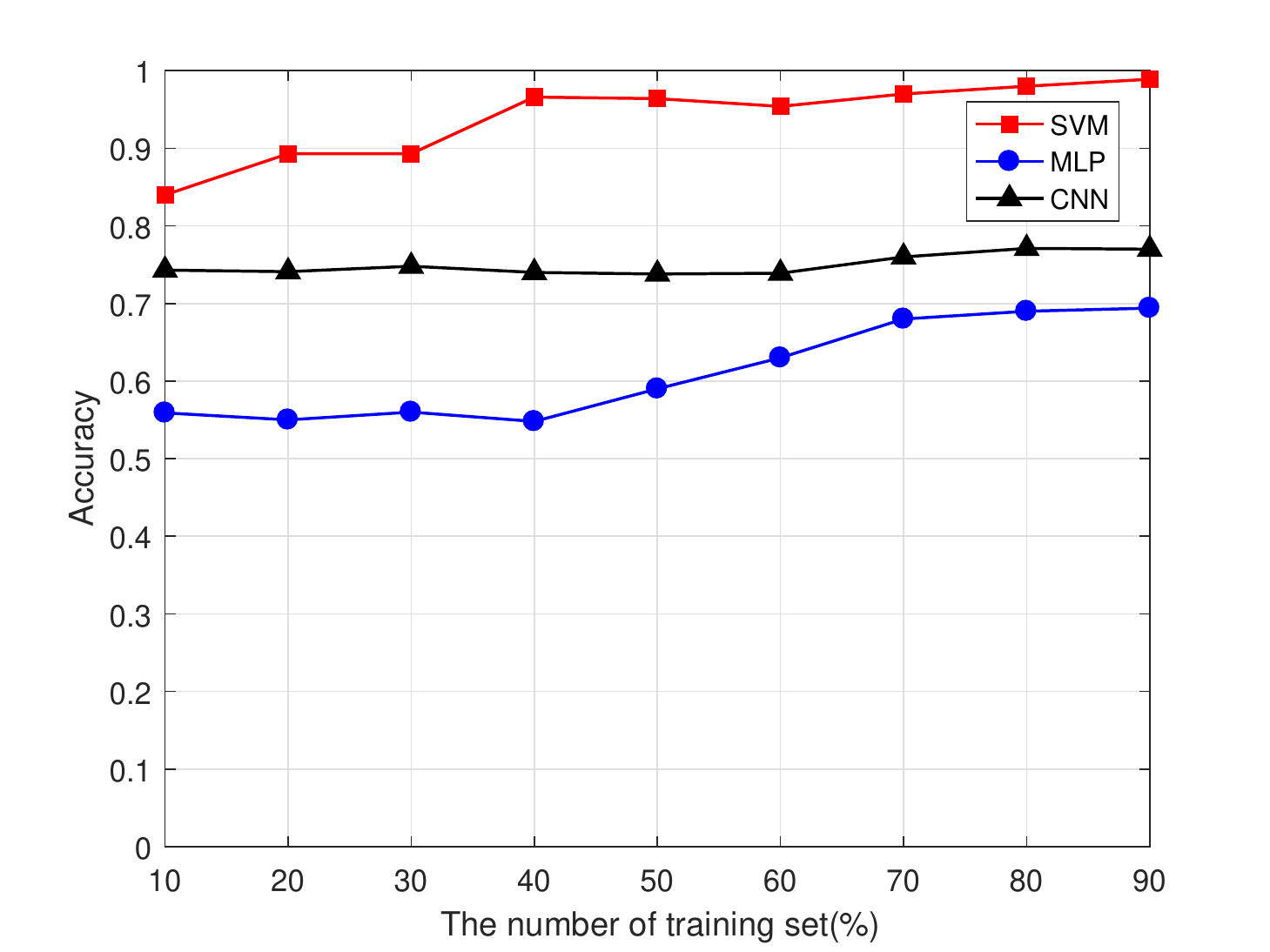}
  \caption{The accuracy of different number of training set.}
  \label{fig:accuracy} 
\end{figure}

\paragraph{\textbf{Comparison 2: training set }}
The neural network structures generally require a large amount of training data to train the model due to the complicated parameters. Whereas, SVM can achieve a high accuracy based on a small training set. We use different proportions of training set to train the classifiers, and the accuracy of the three classifiers are shown in Fig.~\ref{fig:accuracy}. It can be seen that for these three classifiers, the accuracy all increases when the number of training set increases. More specifically, the accuracy of CNN do not improve significantly, which is probably because the amount of training set is not enough for neural network structures. However, SVM can achieve a very high accuracy with a small size of training set.

\paragraph{\textbf{Comparison 3: training time}}
For the classification of images, the training time will generally be affected by the size of images. When comparing the training time of neural networks and traditional classifiers, it is also necessary to consider the number of iterations. In general, SVM requires much fewer iterations than CNN to achieve convergence. For a small training set, the training time of SVM will be much smaller than CNN. As shown in Table~\ref{tab}, it is a comparison of training time for the three classifiers. It can be seen that for the same training data, the training time of the SVM is much smaller than CNN and MLP with the setting of 10 epochs and 64 batch size.

\begin{table}[!htbp]
\centering
\caption{The training-time comparison of classifiers}\label{tab}
\begin{tabular}{|c|c|c|}
\hline
\bfseries Algorithms  &  \bfseries Training Time \textsl{(70\% dataset)} \\
\hline
 \textsl{ \bfseries   Support Vector Machine  }   &  46 s \\
\hline
\textsl{  \bfseries  Convolutional Neural Network   }   & {  474s ($batch\_size=64,epochs=10$)  } \\
\hline
\textsl{\bfseries Multilayer Perceptron  }   &{  105s ($batch\_size=64,epochs=10$)   } \\
\hline
\end{tabular}
\end{table}

\paragraph{\textbf{Comparison 4: test accuracy}}
We randomly split the path image dataset into the training set and the testing set with a ratio of 7:3. We use the balanced training set to train the classifier, and use the testing set in real situation to test the classifiers. The test accuracy comparison of the three classifiers is as shown in Table~\ref{tab5}. It can be seen that CNN is not suitable for the path image classification in our experiment. In contrast, the traditional SVM classifier have an unexpectedly high test accuracy. After our analysis of the classifiers and experimental data, we think it is because there is so little pixel information in path images that CNN model cannot obtain much effective information through the convolution layers. In addition, our training set is not large enough for the neural networks. Therefore, SVM can achieve a higher accuracy than CNN and MLP for the path image classification.

\begin{table}[!htbp]
\centering
\caption{The accuracy comparison of classifiers}\label{tab5}
\begin{tabular}{|c|c|c|}
\hline
\bfseries Algorithms  &  \bfseries Test Accuracy \textsl{(Balanced Training Set)} \\
\hline
 \textsl{ \bfseries   Support Vector Machine  }   &  0.97 \\
\hline
\textsl{  \bfseries  Convolutional Neural Network   }   & 0.76 \\
\hline
\textsl{\bfseries Multilayer Perceptron  }   & 0.68 \\
\hline
\end{tabular}
\end{table}

 In conclusion, although CNN is widely used in image classification, it is not suitable for our dataset. Considering the above four aspects of comparison, SVM is more suitable for path image classification. SVM can not only achieve a high accuracy for a small-sized training set, but also has a short training time and more convenient adjustment of parameters. Therefore, we select the traditional SVM classifier which has a best classification effect on the path image dataset.

\subsubsection{Dependability Evaluation}
\begin{figure}[!htbp]
  \centering
  \subfigure[PR curve for imbalance dataset]{
    \label{fig:cases:a} 
    \includegraphics[width=5.12cm]{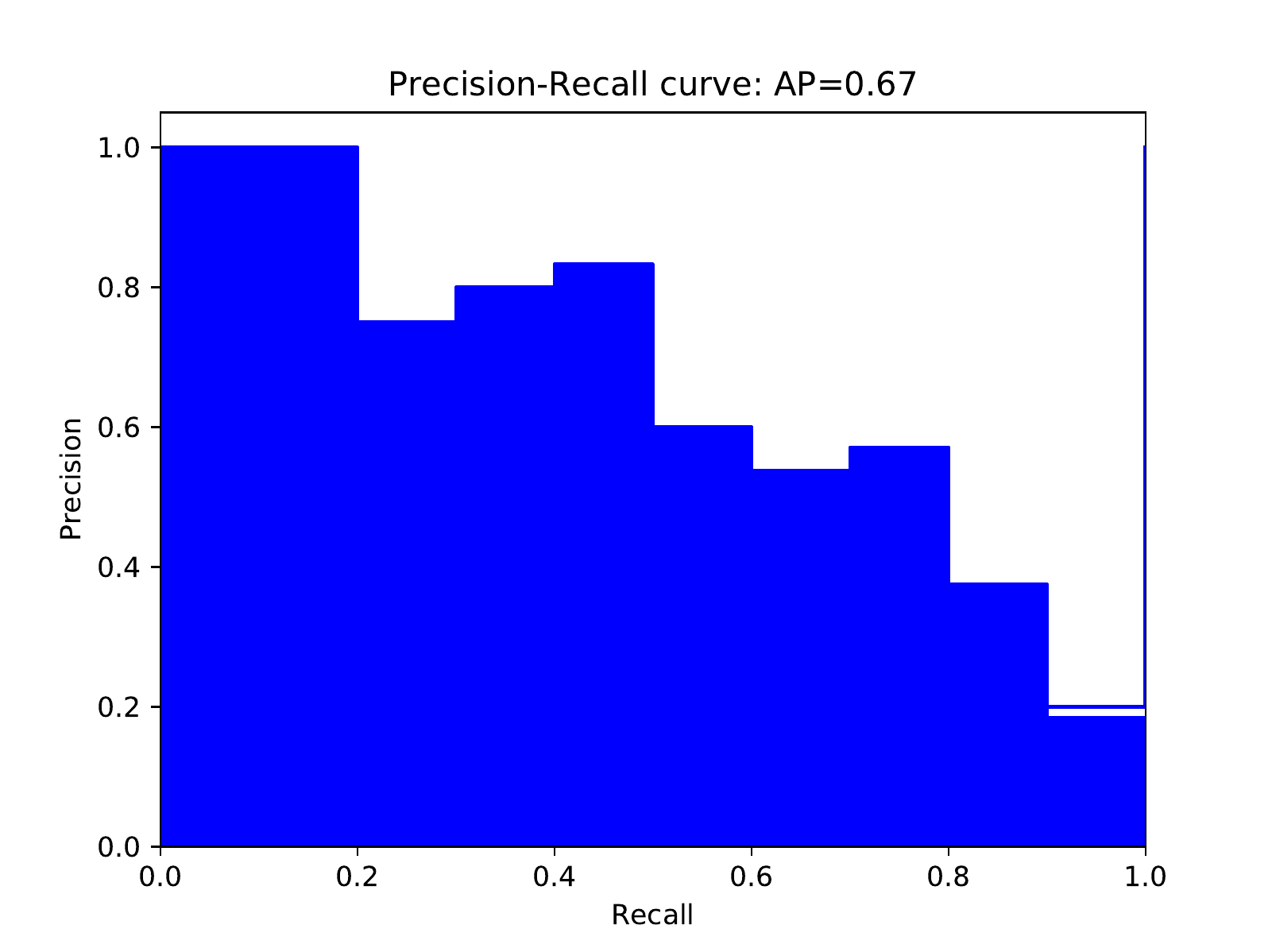}}
  \subfigure[PR curve for balance dataset]{
    \label{fig:cases:b} 
    \includegraphics[width=5.12cm]{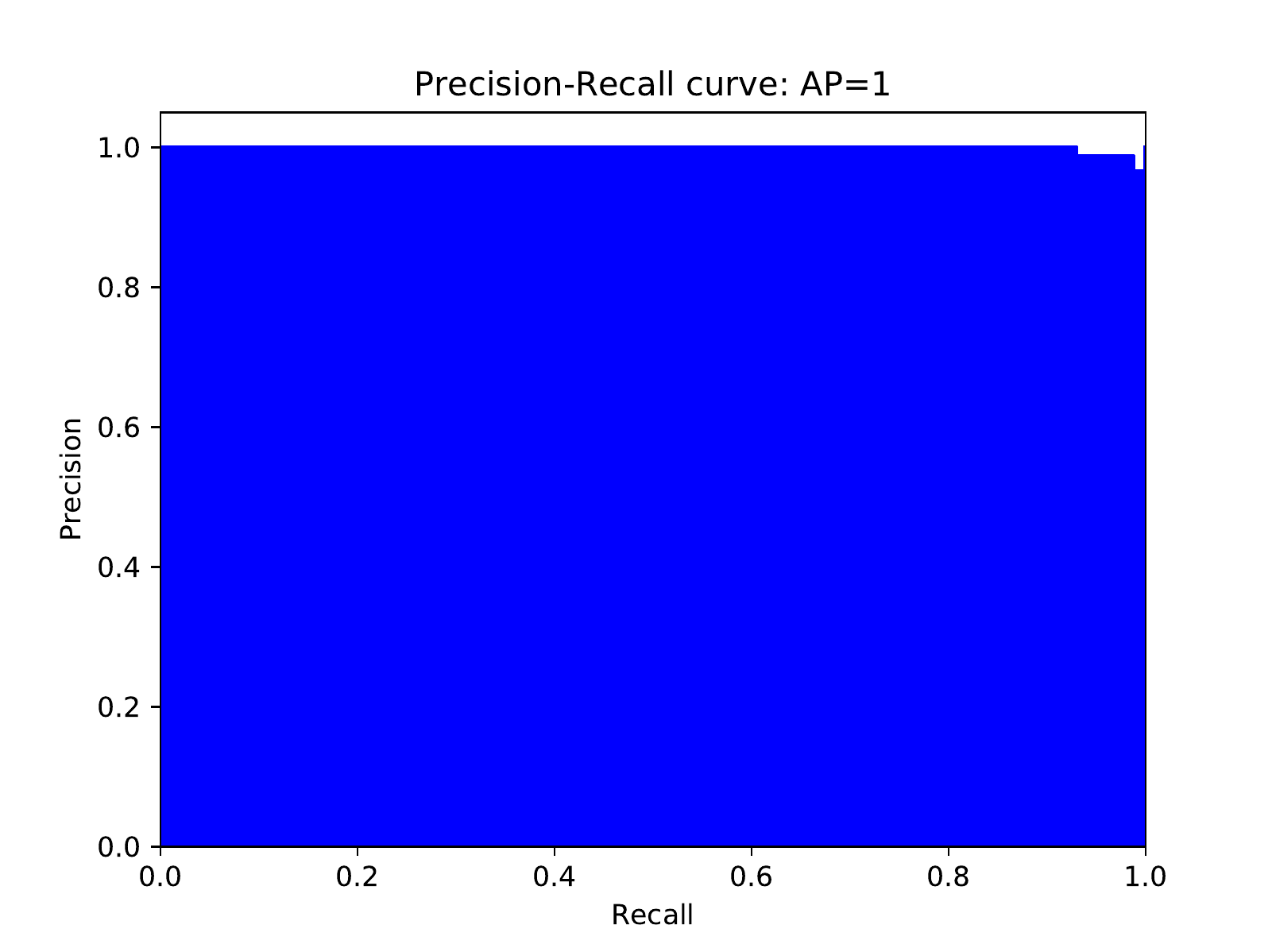}}
    \subfigure[ROC curve for imbalance dataset]{
    \label{fig:cases:c} 
    \includegraphics[width=5.12cm]{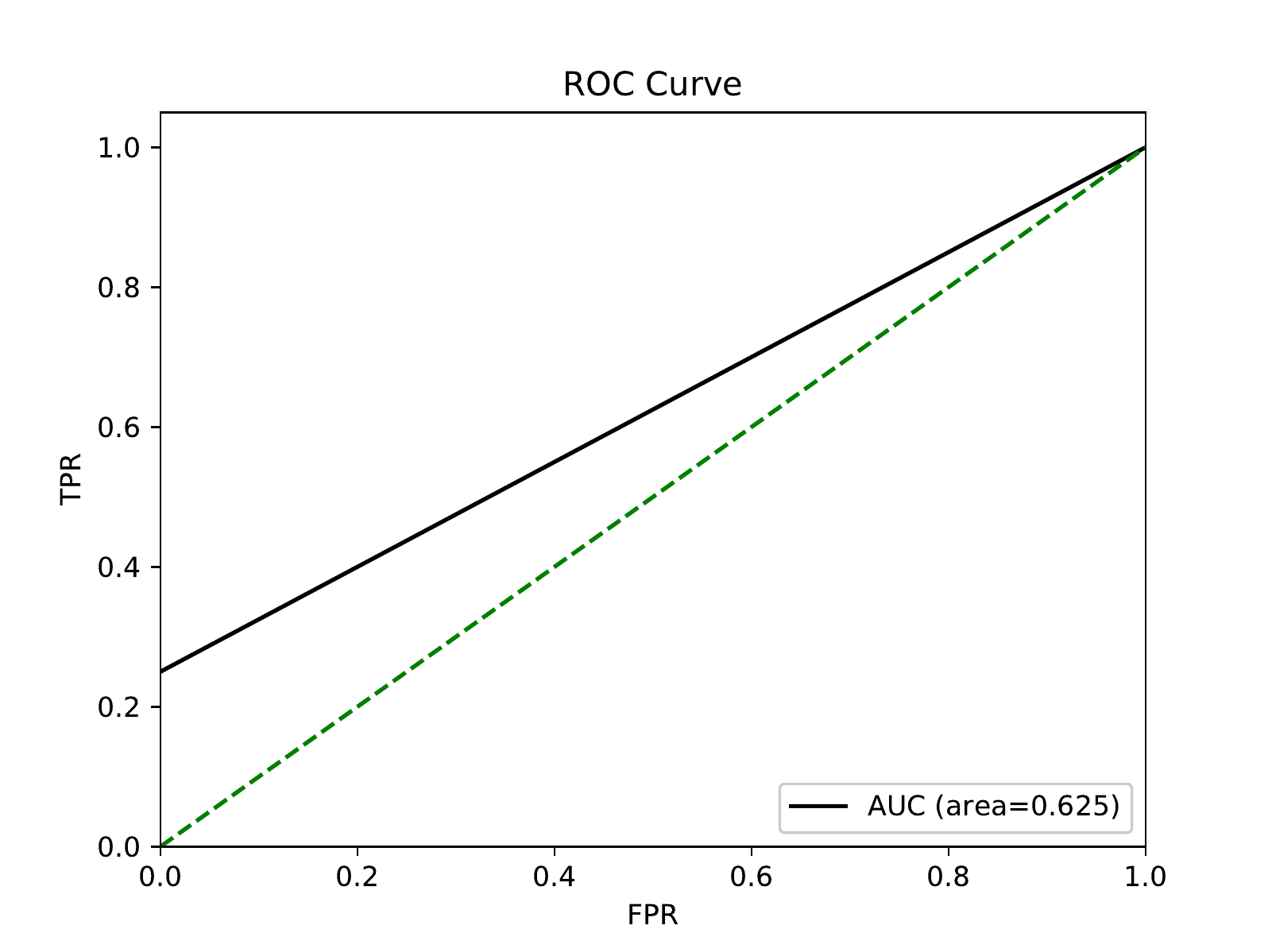}}
      \subfigure[ROC curve for balance dataset]{
    \label{fig:cases:d} 
    \includegraphics[width=5.12cm]{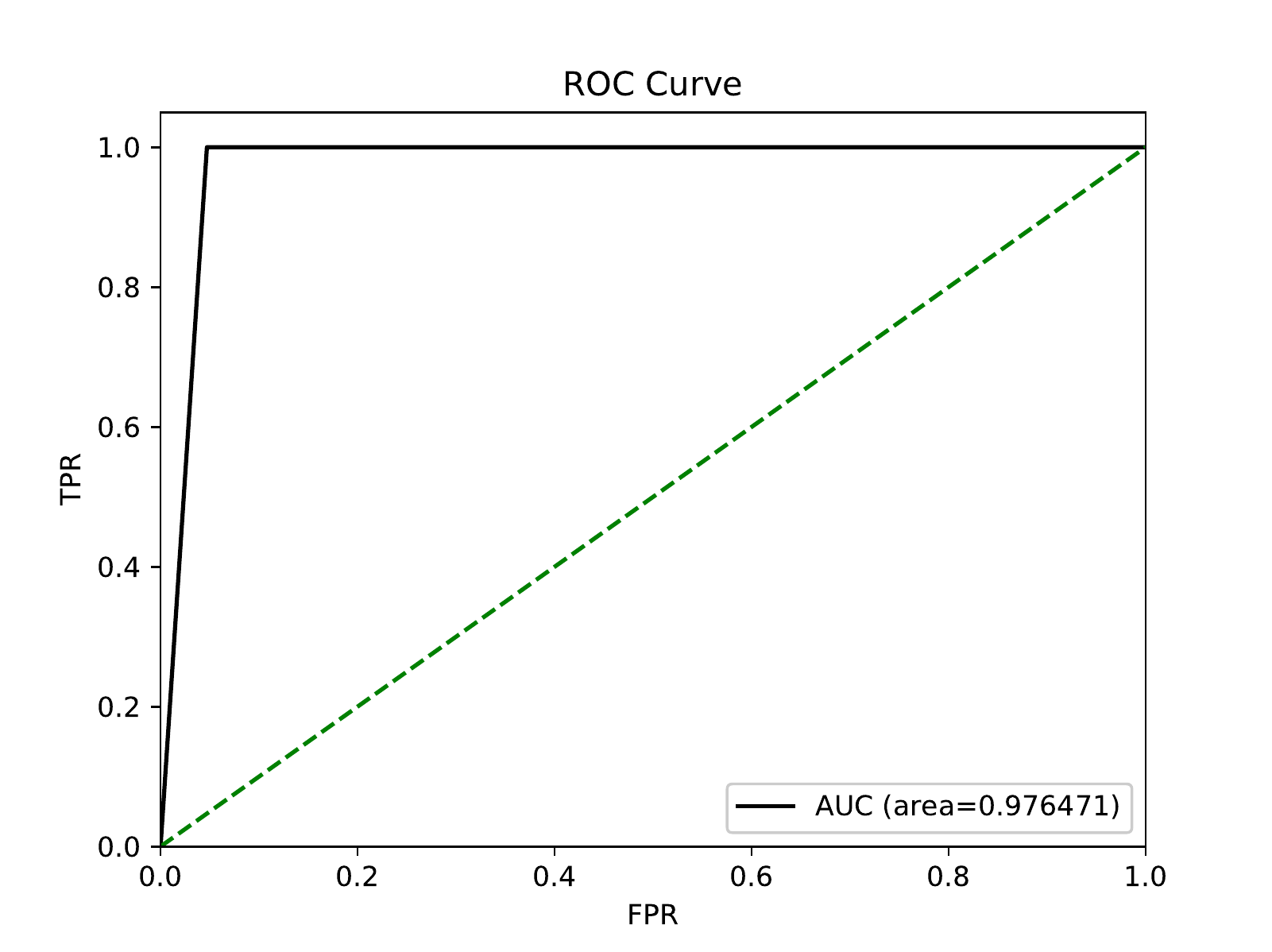}}
  \caption{The PR and ROC curve of SVM before and after data augmentation}
  \label{fig:curves} 
\end{figure}

In addition, we evaluate the confidence of SVM classifier using ROC curve and Precision-Recall curve (see Fig~\ref{fig:curves}). For the imbalance dataset, although the test accuracy reaches 0.91, the AP of PR curve and the AUC of ROC curve are relatively low, indicating that the classification is affected by the categories. After data augmentation, the test accuracy reaches 0.97 and the AP and AUC are close to 1. This illustrates that the test accuracy of SVM classifier is in high confidence.

\subsubsection{Feasibility Evaluation}
We selected 100 path pairs from the dataset to evaluate the processing time, including 40 path pairs from FP and DP categories in the testing set, and 60 path pairs randomly selected from UrP and UcP categories. We compare the processing time of our training-based method and the manual observation method. The processing-time comparison of two methods are shown in the Table~\ref{tab7}. Obviously, the method proposed in this paper has a considerable processing speed and can realize a fast identification of VIN adversarial examples.

\begin{table}[!htbp]
\centering
\caption{The processing-time comparison of two methods}\label{tab7}
\begin{tabular}{cc}
\hline\hline
\bfseries Methods  &   \textsl{ \bfseries   Processing Time (100 path pairs)   }  \\
\hline
 \textsl{\bfseries Training-based Method}         &  0.22s \\
\textsl{\bfseries Manual Observation Method}  & 300s  (\textsl{approximately}) \\
\hline\hline
\end{tabular}
\end{table}

\noindent In summary, the identification approach we proposed can achieve a fast identification for VIN adversarial examples. The combination of path features  comparison and path images classification can ensure the high-accuracy of the classification and the high-speed of the identification approach.
\section{Conclusion}
This paper mainly focuses on the VIN adversarial examples in the domain of robot path planning. We define the four categories of adversarial maps by analysing the possible impacts that adversarial maps may cause on VIN path planning. Based on the categories definition, we implement a training-based identification method by combining the path feature comparison and path images classification. The experiments prove that our method can effectively identify VIN adversarial examples automatically. In future work, we will further study whether our identification approach is applicable to the adversarial examples of other DRL algorithms in the domain of path planning.

\section*{Acknowledgement}
The authors would like to thank the guidance of Professor Wenjia Niu and Professor Jiqiang Liu. Meanwhile this research is supported by the National Natural Science Foundation of China (No.61672092), Science and Technology on Information Assurance Laboratory (No.614200103011711), the Project (No.BMK2017B02-2), Beijing Excellent Talent Training Project, the Fundamental Research Funds for the Central Universities (No.2017RC016), the Foundation of China Scholarship Council, the Natural Science Foundation of China under Grants 61672092, the Fundamental Research Funds for the Central Universities of China under Grants 2018JBZ103.

%
%
%
\bibliographystyle{splncs04}
\bibliography{mybibliography}

\end{document}